\documentclass{article}

\def\be{\begin{equation}}
\def\ee{\end{equation}}
\def\bea{\begin{eqnarray}}
\def\eea{\end{eqnarray}}
\def\({\left(}
\def\){\right)}
\def\<{\left<}
\def\>{\right>}

\def\[{\left[}
\def\]{\right]}
\begin{document}

\pagestyle{empty}
\vskip-10pt
\vskip-10pt
\hfill {\tt hep-th/0411253}
\begin{center}
\vskip 3truecm
{\Large\bf
Dynamics of a wavy plane Wilson surface observable from AdS-CFT correspondence
}\\ 
\vskip 2truecm
{\large\bf
Andreas Gustavsson\footnote{a.r.gustavsson@swipnet.se}
}\\
\vskip 1truecm
{\it Institute of Theoretical  Physics,
Chalmers University of Technology, \\
S-412 96 G\"{o}teborg, Sweden}\\
\end{center}
\vskip 2truecm
\noindent{\bf Abstract:}
Guided by the paper hep-th/0002106 by Polyakov and Rychkov, we compute the second variational derivative of a wavy plane Wilson surface observable, to find that a necessary condition for a proposed surface equation to be satisfied in the large $N$ limit is that we are in the critical dimension $D=6$.
\vfill \vskip4pt

\eject
\newpage
\pagestyle{plain}

\section{Introduction}
To a $k$-dimensional submanifold $\Sigma_k$ embedded in $D$-dimensional space one may sometimes associate an observable $W(\Sigma_k)$. For $k=1$ this is a Wilson loop. Higher $k$'s arise whenever one has a charged $(k-1)$-brane, to whose world-sheet there is associated a $k$-submanifold observable. Non-abelian Wilson loops can be defined using a path ordering operation which is invariant under reparametrizations of the loop. For surfaces we can not define such a `surface ordering' in an invariant way \cite{Teitelboim}, unless we also introduce a one form connection which enable us to define parallel transportation along curves on the surface \cite{Alvarez}. But we will not use this approach here to compute the non-abelian Wilson surface. 

We will here investigate the situation where a two-form potential couples to a selfdual string in an M five-brane. That is, the case when $k=2$. The associated observable is then usually called a Wilson surface. The theory living on a stack of $N$ five-branes is associated with the $A_{N-1}$ Lie algebra, and it is natural to call the corresponding theory for $A_{N-1}$ theory. These theories are really completely characterized by giving its associated Lie algbra. They have no other free parameters. Otherwise very little is known about these theories. Though a definition of these theories was suggested in \cite{Ganor}, in terms of a `surface equation'. We will in this letter examine whether such a surface equation can be satisfied by the realization of the Wilson surface in the AdS-CFT correspondence. If that could be shown, that would put the surface equation on a more solid ground. So far the surface equation has only been guessed, and has not been derived (other than in the abelian case). This is to be compared with the loop equation, which has been derived from Yang-Mills theory \cite{Polyakov}\cite{Ooguri}\cite{Migdal} (a review is \cite{Makeenko}). Showing that the loop equation is satisfied by its realization as a string in AdS space would prove the AdS-CFT correspondence, and such a derivation was initiated by Polyakov and Rychkov \cite{Polyakov-Rychkov}\cite{Polyakov-Rychkovb}. But they did not treat explicitly the really interesting cases when one has a cusp or self-intersection on the loop. We will of course not do it better here, with Wilson surfaces instead, but will just consider a smooth surface without self-intersections, being close to a flat plane. That is, the simplest non-trivial case.

According to the AdS-CFT correspondence the expectation value of a submanifold observable should be given by \cite{Maldacena}\cite{Yang-Yee}
\be
\<W(\Sigma_k)\>\sim\exp-TA(\Sigma_k)
\ee
Here $T$ is the brane tension of a $(k+1)$-dimensional brane wrapping a minimal surface $M_{k+1}$ in AdS-space whose boundary is $\Sigma_k$, and $A(\Sigma_k)$ is the area of $M_{k+1}$ (the minimal surface $M_{k+1}$ is usually uniquely determined by its boundary $\Sigma_k$). This formula is not quite correct since we should somehow sum over all surfaces in AdS space whose boundary are the surface observable. But in a classical limit (large $N$ limit) this sum can be approximated using steepest descent, which is to evaluate the area at its extremum point, which is at the minimal surface. However the area of this minimal surface is infinite. The common procedure to regularize it is to let the minimal surface instead end at $y=\epsilon$ in Poincare coordinates (in which the metric is $ds^2=y^{-2}dy^2+...$) and not directly at the boundary at $y=0$. When one does such a regularization for $k=2$, one finds \cite{Graham-Witten}
\bea
A(\Sigma_2)=B\epsilon^{-2}+C\ln\epsilon+D\label{Willmore}
\eea
where $2B$ is the area of the surface, and $8C$ is the rigid string action, also called the Willmore functional. $D$ is the renormalized Wilson surface. Due to the logarithmic divergence, a rescaling of $\epsilon$ by any real dimensionless number $e^{a}>0$ gets transmuted into a shift $D\rightarrow D+aC$ of the renormalized Wilson surface. $C$ is a scaling anomaly.

In this paper we will mainly concentrate on the case that $k=2$, that is on Wilson surfaces. We will work in euclidean signature. Following Polyakov and Rychkov \cite{Polyakov-Rychkov} we will (at least in principle) compute the Wilson surface for a surface that is a fluctuating infinitely extended plane in the large $N$ limit of $A_N$ theory, by realizing the Wilson surface as a minimal surface in AdS space. Apart from being a situation where we can make explicit computations, the wavy plane might also be of wider interest, being an observable associated with a wavy selfdual string. Though we work in Euclidean signature, Wick rotation to Minkowski signature should be permissible since we work in a flat background.

We then compute the second variational derivative of this Wilson surface. Our aim with that computation is to check whether a surface equation like
\bea
L(\sigma)\<W\>=?
\eea
really could hold in the large $N$ limit. The surface operator, that we denote by $L$, is a kind of Laplace operator on the space of surfaces. It is a direct generalization of the Polyakov loop operator \cite{Polyakov} and is defined by Eq. (\ref{L}). We will find that a necessary condition for the surface equation to be satisfied is that we take the number of dimensions to be $D=6$.

A rather similar surface equation was proposed in \cite{Ganor}, where it also was given a suggestion of what the right-hand side of the surface equation would be for the case that the surface self-intersects. 

In appendix $C$ we show that the surface operator, generalized to $k$-submanifolds, is conformally invariant precisely in $D=2k+2$ dimensions. This generalizes the proof given in \cite{Polyakov-Rychkov} for the loop.

\section{The abelian surface equation}
In abelian theory the Wilson surface is given by
\be
W(\Sigma)=\exp{i\int_{\Sigma}B}
\ee
where $B$ denotes a two-form connection. We denote its curvature by $H=dB$. From Stokes theorem it follows that $W$ is invariant under local gauge transformations $B\rightarrow B+d\lambda$, meaning that $W$ is an observable. Globally we have that $W$ is invariant under $B\rightarrow B+\Lambda$ provided $d\Lambda=0$ and $\int_{\Sigma} \Lambda\equiv 0$ mod $2\pi$. For this to hold, the kinetic term should be normalized as $\frac{1}{8\pi}\int H\wedge*H$ \cite{Henningson}.

We would like to see if we could define a surface operator as some kind of Laplace operator in surface space (the space of surfaces). We compute the second variational derivative of the abelian Wilson surface. Parametrizing the surface $\Sigma$ as $\sigma^i\mapsto X^{\mu}(\sigma)$, ($i=0,1$, $\mu=0,1,2,3,4,5$) we have
\bea
\frac{\delta^2 \<W\>}{\delta X^{\mu}(\sigma)\delta X^{\mu}(\sigma')}&=&i\<\partial_{\mu} H_{\mu\nu\rho}(X(\sigma))W\>\dot{X}^{\nu}(\sigma)X'^{\rho}(\sigma)\delta^2(\sigma-\sigma')\cr
&&-\<H_{\mu\nu\rho}(X(\sigma))H_{\mu\kappa\tau}(X(\sigma'))W\>\dot{X}^{\nu}(\sigma)X'^{\rho}(\sigma)\dot{X}^{\kappa}(\sigma')X'^{\tau}(\sigma').\label{second1}
\eea
We define the surface operator $L(\sigma)$ as
\be
\frac{\delta^2 \<W\>}{\delta X^{\mu}(\sigma)\delta X^{\mu}(\sigma')}\equiv L(\sigma)\<W\>\delta^2(\sigma-\sigma')+...\label{L}
\ee
where $...$ are terms that do not involve delta function singularities. From (\ref{second1}) we then see that $L(\sigma)$ has the effect of bringing down the equation of motion for the gauge field. Classically the right-hand side (in the absence of charged strings) is zero. In quantum theory we get a contact term
\be
\frac{1}{4\pi}\<\partial_{\mu} H_{\mu\nu\rho}(x)W\>=iJ_{\nu\rho}(x)\<W\>.
\ee
Here $J_{\mu\nu}(x)=(*\delta)_{\mu\nu}(x)$ and $\delta$ denotes the Poincare dual of $\Sigma$ and $*$ the Hodge duality operator. Since in the abelian theory, the operator product in the second term in the right-hand side of (\ref{second1}) does not contain any delta function singularity, we get the abelian surface equation
\bea
L(\sigma)\<W\>=4\pi J(X(\sigma))\<W\>,
\eea
where
\bea
J(X(\sigma))\equiv \int_{\Sigma}d^2\sigma' \dot{X}_{\mu}(\sigma)X'_{\nu}(\sigma)\dot{X}_{[\mu}(\sigma')X'_{\nu]}(\sigma')\delta^6(X(\sigma)-X(\sigma')).\label{J}
\eea
But this $J$ has to be defined, since naively it is infinite. To find out how to define $J$ we must first choose a regularization prescription of $\<W\>$. If we for instance use a cutoff regularization in momentum space defined such as that it maps a function (which may or may not be singular at $x=0$) into a regularized function according to
\bea
f(x)\mapsto f_{\epsilon}(x)=\int \frac{d^6 p}{(2\pi)^6}\int d^6 y e^{ip.(x-y)}e^{-(\epsilon |p|)^2}f(y),
\eea
then we find that the delta function will get mapped into a regularized delta function as follows,
\bea
\delta^6(x)\mapsto \delta_{\epsilon}^6(x) = \frac{1}{64\pi^3\epsilon^6}e^{-\frac{|x|^2}{4\epsilon^2}}.
\eea
Using cutoff regularization, we should thus define
\be
\delta^6(X(\sigma)-X(\sigma'))\equiv \frac{1}{64\pi^3}\lim _{\epsilon\rightarrow 0}\frac{1}{\epsilon^{6}}\exp\(-\frac{|X(\sigma)-X(\sigma')|^2}{4\epsilon^2}\).
\ee
To see what this definition really means, we have to integrate it with a test function $u(\sigma)$ as
\bea
\int d^2\sigma u(\sigma)  \delta^6(X(\sigma)-X(\sigma')).\label{test}
\eea
If the surface is smooth and has no self-intersection, we may expand $X(\sigma)$ around $\sigma'$, and also make a similar expansion in powers of $\rho\equiv \sigma-\sigma'$ of $f(\sigma)$. We get a finite contribution only from the fourth order term in $\rho$,
\bea
\int d^2\rho \rho^4 \frac{1}{64\pi^3}\epsilon^{-6} e^{-\rho^2/\(4\epsilon^2\)}=2\pi^{-2}.
\eea
We could instead use dimensional regularization and should get the same finite result. In dimensional regularization we should arrive at the following definition of  the delta function \cite{Polyakov}
\bea
\delta^6(X(\sigma)-X(\sigma'))\equiv\frac{1}{\pi^3}\lim_{D\rightarrow 6}\frac{6-D}{|X(\sigma)-X(\sigma')|^D}.
\eea
Here $\pi^3$ is the volume of the unit five-sphere. Using this definition we again find that only the terms at fourth order in $\rho$ gives a non-zero contribution to (\ref{test}). This time because 
\be
\lim_{D\rightarrow 6} \frac{1}{\pi^3}(6-D)\int d^2\rho \rho^{4-D}=2\pi^{-2}.
\ee
We see that we get exactly the same numerical result for both these regularizations. For a flat plane parametrized as $X^i(\sigma)=\sigma^i$, $X^{i'}(\sigma)=0$ ($i=0,1$, $i'=2,3,4,5$) we only get a contribution at fourth order from expanding the test function $u(\sigma)=u(\sigma')+...+(1/24)\rho^i\rho^j\rho^k\rho^l\partial_i\partial_j\partial_k\partial_lu(\sigma')+...$. We then find that
\bea
\delta^6(X(\sigma)-X(\sigma'))=\frac{1}{32\pi^2}(\partial_i\partial_i)^2\delta^2(\sigma-\sigma').
\eea
If the surface is not a flat plane, we also get contributions from expanding $|X(\sigma)-X(\sigma')|$ in powers of $\rho$. To find out the meaning of $\delta^6(X(\sigma)-X(\sigma'))$ in this case, we consider integrating it with a test function
\bea
\int d^2\sigma u(\sigma)\frac{1}{|X(\sigma)-X(\sigma')|^D}.
\eea
Expanding $|X(\sigma)-X(\sigma')|^2$ and keeping only the sixth order contribution in $\rho$ ,we find in this way\footnote{This is just one of many regularizations one may choose. We have checked that a cut-off regularization yields the same result.} that for a surface being a graph $X^{\mu}(\sigma)=(\sigma^i,\phi^{i'}(\sigma))$, the delta function is given by
\bea
\delta^6(X(\sigma)-X(\sigma'))&=&\frac{1}{32\pi^2}\bigg(\partial^4\delta^2(\sigma-\sigma')-\frac{1}{6}{\partial_{\sigma}}^6\Big\{(\phi(\sigma)-\phi(\sigma'))^2\delta^2(\sigma-\sigma')\Big\}\bigg)+O(|\phi|^4)
\eea
and 
\bea
J(X(\sigma))=-\frac{1}{32\pi^2}\(\frac{1}{6} \Big(\partial^6(\phi(\sigma).\phi(\sigma))-2\phi(\sigma).\partial^6 \phi(\sigma)\Big)
-\partial_i\phi(\sigma).\partial_i\partial^4\phi(\sigma)\)+O(|\phi|^4).\label{JJ}
\eea
But this is not zero! Not even if we integrate this over $\sigma$ do we get zero. This is very different from the situation one has for Wilson loops where the corresponding delta function $\delta^4(X(\sigma)-X(\sigma))=0$ for smooth loops without self-intersections. Let us review how this comes about for Wilson loops \cite{Polyakov}. Then when we integrate with a test function $u(\sigma)$ we have to compute
\bea
\int d\sigma u(\sigma)\frac{1}{|X(\sigma)-X(\sigma')|^D}
\eea
but here we get a contribution only from the third order terms in $\rho$ because
\bea
\int d\rho |\rho|^{-D}|\rho|^3=\frac{1}{4-D}.
\eea
If the loop has a well-defined tangent vector everywhere, we can, upon expanding everything in $\rho$, only get contributions of the form
\bea
\int d\rho |\rho|^{-D}\rho^3=0.
\eea

Throughout this paper we will use the surface equation as written in momentum space. We first define 
\bea
\frac{\delta^2}{\delta X(k).\delta X(k')}\equiv \frac{1}{(2\pi)^4}\int d^2\sigma\int d^2\sigma' e^{-i(k.\sigma+k'.\sigma')}\frac{\delta^2}{\delta X(\sigma).\delta X(\sigma')}.
\eea
and then
\bea
\tilde{L}(k+k')=\frac{1}{(2\pi)^4}\int d^2\sigma e^{-i(k+k').\sigma} L(\sigma).\label{SurfaceOp}
\eea
As an illustration of this, we may Fourier transform (\ref{JJ}) and obtain the following abelian surface equation in momentum space representation:
\bea
\tilde{L}(q=0)\<W\>=\frac{1}{6\pi}\frac{1}{(2\pi)^4}\int \frac{d^2 p}{(2\pi)^2}\phi(p).\phi(-p)p^6\<W\>+O(|\phi|^4)\label{se}
\eea
where we have taken $q\equiv k+k'=0$ for simplicity. 

The definition of the surface operator (\ref{L}) should lead to a corresponding definition in momentum space representation. This definition should be
\bea
\frac{\delta^2\<W\>}{\delta X(k).\delta X(k')}=\tilde{L}(k+k')\<W\> + o(1)\label{central}
\eea
where we should take $k,k'\rightarrow \infty$. Here we must restrict the piece $o(1)\supset O(k)$ in the large $k$ limit to contain only non-constant terms (a constant term in momentum space would correspond to $\delta^2(\sigma-\sigma')$ in position space). But it is complete nonsence  to separate a generic function into a constant piece and a non-constant piece, since we may just as well define a new non-constant function to be any constant plus the original non-constant function! So we must be more precise in what we mean by non-constant terms. What we will find both in the large $N$ limit as well as for $N=1$, is that the piece that we denoted $o(1)$ will be of the form $kP(k)+kQ(k)\ln|k|$ in the large $k$ limit. Here $P$ and $Q$ are polynomials in $k$. In particular we will find is that odd powers of $k$ necessarily involve at least one vector $k_i$ (which is always contracted with another momentum vector $p_i$ over which we integrate), and we do not get any norms $|k|$ rised to odd powers. Unfortunately that means that we have to worry about whether $kP(k)$ could correspond to a bunch of derivatives acting on $\delta^2(\sigma-\sigma')$. We saw that in the abelian case that no such derivatives did occur, and we would like to think that this situation prevails in any of the $A_{N-1}$ theories as well, but we have no proof. But we think it is plausible that the piece $kP(k)$ is simply an effect of the indeterminacy of the logarithms (this will be explained in a moment). We will also see that this really is the case in an effective $A_1$ model where certainly no derivatives of $\delta^2(\sigma-\sigma')$ arise in the second variational derivative, but nevertheless we find a piece which is of the form $kP(k)$ in the second variational derivative. 

At least we could find an argument that excludes the annoying term $\sim\ln|k|$ in the large $k$ limit, and this was the real reason for why we chose to denote these term by $o(1)$ -- they must all tend to zero as $k\rightarrow 0$. This term would have been very annoying because, due to the indeterminacy of the logarithm, such a term would interfere with the constant piece, that is, with the right-hand side of the surface equation. That would make it impossible to make sense of the surface equation in momentum space. Now taking the limit $k\rightarrow 0$ corresponds to taking a constant mode. Varying the Wilson surface with respect to a constant mode is to study its variation under a rigid translation. Assuming translational invariance, such a variation vanishes. Hence the second variational derivative vanishes in the limit $k,k'\rightarrow 0$.  Hence the term $\ln|k|$ is not allowed in the small $k$ limit since it does not vanish in that limit. Now we may have a function that converges to $\ln|k|$ only in the limit $k\rightarrow \infty$, such as $\ln|p+k|$ for some finite $p$. But not this term either converges to zero in the limit $k\rightarrow 0$ unless possibly for some certain value of $p$, and therefore a term like $\ln|p+k|$ is not allowed either. But this argument can not exclude the possibility of having a term which asymptotically behaves like $\ln|k|$ for large $k$ and which tends to zero as $k\rightarrow 0$, but we consider the possibility of having such a term as very unlikely. The computations that we present in this letter show that the only terms that occur (both in the large $k$ and the small $k$ limits) are polynomials in $k$ and polynomials times a single logarithm.

Using high frequencies $k$ and $k'$, we are able to probe finer details on the surface. To illustrate this important point, we may take a function $f(\sigma)=a\delta^2(\sigma)+b\epsilon^{-2}e^{-\sigma^2/(4\epsilon^2)}$ composed of a sharp delta function and a smeared out delta function, with some associated numerical coefficients $a$ and $b$. In position space we can extract the coefficient $a$ by integrating $f(\sigma)$ as follows, $a=\lim_{\epsilon\rightarrow 0} \int_{|\sigma|<\epsilon}d^2\sigma f(\sigma)$. We now ask what the corresponding operation would be in the momentum space. In momentum space we have $f(k)=a+be^{-\epsilon^2 k^2}$. We see that in order to see the coefficient $a$ of the delta function singularity clearly, that is without also seeing the smeared out delta function, we should take $|k|>>\epsilon^{-1}$, because we have $a=\lim_{k\rightarrow \infty} f(k)$. Now what we will encounter in the second variational derivative will not be gaussian bump functions, but rather something more like $f(\sigma)=a\delta^2(\sigma)+b|\sigma|^{-2}$. In momentum space this becomes $f(k)=a+b\ln(|k|\epsilon)$ where $\epsilon$ is a cutoff. Now a rescaling of $\epsilon\rightarrow e^c \epsilon$ amounts to a term $bc$, and in general it is not possible to distinguish this term from $a$. This was what we refered to as the indeterminacy of the logarithm -- in the renormalized result it looks like the logarithm has been shifted by a constant $c$. 

In the rest of this paper we will take $k'=-k$, for simplicity. Then the definition (\ref{central}) implies a definition of $\tilde{L}(0)\<W\>$, which is manifestly rotationally invariant (The vector $0$ contains no information of directions). To extract $\tilde{L}(0)\<W\>$,  Eq. (\ref{central}) leads us to the following recipe: First compute the second variational derivative of $\<W\>$ when written in momentum space. Then take the limit $k\rightarrow \infty$, this in order to probe the delta function $\delta^2(\sigma-\sigma')$. We should in that result only consider the $k$-independent terms (if one believes that derivatives of $\delta^2(\sigma-\sigma')$ are non-present), and finally we should average the remaining terms over all directions in parameter space.

In the next section we will compute the Wilson surface (and it second variational derivative) in effective $A_1$ theory for the case of a wavy plane, parametrized as
\bea
X^i(\sigma)&=&\sigma^i\cr
X^{i'}(\sigma)&=&\phi^{i'}(\sigma)\label{wavy}
\eea
to a certain order in its transverse fluctuations $\phi^{i'}$. But before we embark on that computation, let us first see if we can compute something directly from the right-hand side of (\ref{second1}), to lowest order in the $\phi$'s. Consider the second term in the right-hand side. Here we need the operator product expansion of
\bea
&&H_{\mu\nu\rho}(X(\sigma))H_{\mu\tau\sigma}(X(\sigma'))\dot{X}_{\nu}(\sigma)X'_{\rho}(\sigma) \dot{X}_{\tau}(\sigma')X'_{\sigma}(\sigma')\cr
&&=H_{i'01}(X(\sigma))H_{i'01}(X(\sigma'))\cr
&&+H_{i'j'1}(X(\sigma))H_{i'k'1}(X(\sigma))\dot{\phi}^{i'}(\sigma).\dot{\phi}^{k'}(\sigma')
+H_{i'0j'}(X(\sigma))H_{i'0k'}(X(\sigma'))\phi'^{i'}.\phi'^{k'}\cr
&&+O(\phi^4).\label{ope}
\eea
To compute this carefully, we would need to take into account all non-singular terms in the OPE. We will not do that this carefully (since we will make this computation in a different way later), but just consider the first few terms, 
\bea
H_{\mu\nu\rho}(x)H^{\kappa\tau\sigma}(y)=9\int \frac{d^6k}{(2\pi)^6}\frac{16\pi}{|k|^2}\delta_{[\mu\nu}^{[\kappa\tau}k_{\rho]}k^{\sigma]}e^{ik.(x-y)}+...,
\eea
written in momentum representation. We will find it convenient to define the following functions\footnote{Here we display only the renormalized results. We could for instance insert a cutoff $e^{-\epsilon\sqrt{p^2+P^2}}$ in the divergent integrals, or we could use dimensional regularization, to get at these results. Furthermore $\ln|p|$ can only be defined up to an additiv constant that depends on the renormalization scheme. In cutoff regularization, we get a divergence $\ln(\epsilon|p|)$ (We also get a divergence $\sim\epsilon^{-2}$). A rescaling of $\epsilon$ amounts to a shift of the logarithm. In dimensional regularization a rescaling of momenta leaves the integrals unchanged, but the effect of such a rescaling on the logarithm $\ln|p|$ is again to shift it by a dimensionless constant.} 
\bea
H(p)&=&\int d^4P\frac{1}{p^2+P^2}=V_3\(-\frac{p^2}{2}+p^2\ln|p|\)\cr
G(p)&=&\int d^4P\frac{P^2}{p^2+P^2}=V_3\(\frac{3}{4}p^4-p^4\ln|p|\)\cr
F(p)&=&\int d^4P\frac{P^4}{p^2+P^2}=V_3\(-\frac{7}{12}p^6+p^6\ln|p|\)\label{fns}
\eea
where $V_3=2\pi^2$ denotes volume of unit three-sphere. To examine the contribution coming from, say, the variational derivative $\frac{\delta^2}{\delta \phi^{i'}\delta \phi^{i'}}$ we need only consider the first term $H_{i'01}H_{i'01}$, because the next term will be of fourth order. Thus we compute
\bea
&&\int d^2\sigma\int d^2\sigma' e^{ik.(\sigma-\sigma')} H_{i'01}(X(\sigma))H_{i'01}(X(\sigma'))\cr
&=&\frac{16\pi}{(2\pi)^4}\Bigg[(2\pi)^2\delta^2(0)\(\frac{1}{2}G(k)+2k^2H(k)\)\cr
&&-\int \frac{d^2 p}{(2\pi)^2} \phi(p).\phi(-p)\(\frac{1}{2}(F(k)-F(k+p))+2k^2G(k)-2(k+p)^2G(k+p)\)\Bigg]
\eea
or explicitly,
\bea
&=&V_3\frac{16\pi}{(2\pi)^4}\Bigg[(2\pi)^2\delta^2(0)\(-\frac{5}{8}k^4+\frac{3}{2}k^4\ln|k|\)\cr
&&-\int \frac{d^2 p}{(2\pi)^2}\phi(p).\phi(-p)\(\frac{29}{24}\(k^6-(k+p)^6\)+\frac{3}{2}\((k+p)^6\ln|k+p|-k^6\ln|k|\)\)\Bigg].
\eea
This representation is not unique because we may shift the logarithms by a constant (see the footnote). The terms which are powers in momenta look precisely like terms that just as well could have arisen from having Fourier transformed delta function singularities or derivatives thereof. But this is of course not the case here. Here they arose as a consequence of the indeterminacy of the logarithms.

\section{A wavy plane in low energy effective $A_1$ theory}\label{wp}
Due to our lack of knowledge in the theory living on many parallel M five-branes and in how to compute expectation values of non-abelian Wilson surfaces, we will here restrict ourselves to just two separated five branes. At low energies (compared to their separation) the theory is accurately described by an effective $A_1$ theory in terms of the field content in just one tensormultiplet. In \cite{Gustavsson} we showed that the Wilson surface to be used in the context of AdS-CFT should be\footnote{In a supersymmetric theory one should include fermions in this definition, in line with \cite{Ooguri}.}
\be
W=e^{i\int \(B+d^2\sigma\sqrt{g}\varphi\)}\label{surface}
\ee
where $\varphi$ is a scalar field which has vacuum expectation value zero. One motivation for this definition is that with this definition one gets a conformal anomaly which is proportional to the conformal anomaly one derives in the large $N$ limit\cite{Gustavsson}. If one is careful with normalizations one can also show that the proportionality constant is precisely equal to $N$. So it seems as the 'non-Abelian' Wilson surface would in some sense contain a product of $N$ Wilson surfaces of this type. A more precise way to understand this is by separating $N$ M-theory five branes from each other. One then expects to get a theory of $N$ copies of free tensor multiplets\footnote{This is not entirely true, as was shown in \cite{Intriligator}, no matter how far we separate our five branes. Also notice that we should use the Wilson surface with the scalar $\varphi$ rather than the abelian Wilson surface that we would use if there was just one five brane}, and correspondingly the Wilson surface should be a product of $N$ copies of Wilson surfaces as defined in (\ref{surface}). This gives an intuitive explanation why we get a conformal anomaly which is proportional to the large $N$ limit with proportionality constant $N$. We would not expect this anomaly to change in a process where we continuosly separate the five branes.

We now set out to compute the wavy plane for the effective $A_1$ theory. Our purpose with this computation is to demonstrate (\ref{central}) in an explicit computation. We admit that this computation is somewhat tautological of what we have already done. But we want to compute the second variational derivative of the wavy plane in detail, so that we can compare this with the corresponding computation that we will do in the large $N$ limit, and the safest way to do that seems to be by first computing the Wilson surface, and then computing its second variational derivative. But in principle one should be able to make the computation of the operator product and of the delta function in detail and arrive at the same result. Though the computation we present here requires no knowledge of operator product expansions, nor of tricky delta functions on surfaces. These concepts will be implicit in this computation.

The vacuum expectation value of the Wilson surface is given by $\<W(\Sigma)\>=\exp -S(\Sigma)$ with $S=S_B+S_{\varphi}$ where
\be
S_B(\Sigma)=\int dX^{\mu}\wedge dX^{\nu}\int dY^{\kappa}\wedge dY^{\tau}D_{\mu\nu,\kappa\tau}(X-Y),
\ee
and where the gauge field propagator is given by
\be
D_{\mu\nu}^{\kappa\tau}(x-y)=\int \frac{d^6 p}{(2\pi)^6} e^{ip.(x-y)}\frac{16\pi}{p^2}\(\delta_{\mu\nu}^{\kappa\tau}+\xi \frac{\delta_{[\mu}^{[\kappa}p_{\nu]}p^{\tau]}}{p^2}\).
\ee
Physical quantities must not depend on the gauge parameter $\xi$. This is to say that our regularization must preserve the identity
\bea
0&=&d^6p\int dX^{\mu}\wedge dX^{\nu}\int dY_{\rho}\wedge dY_{\nu}\frac{p_{\mu}p^{\rho}}{p^4}e^{ip.(X-Y)}\cr
&&=-\int d^6p\frac{1}{p^4}\int_{\Sigma} d\[dX^{\nu}e^{ip.X}\]\int_{\Sigma} d\[dY_{\nu}e^{-ip.Y}\]
\eea
which formally holds due to Stokes theorem for a closed surface $\Sigma$. We see that we can regularize by inserting the factor $e^{-\epsilon |p|}$, or use dimensional regularization, without altering this identity. We thus consider
\be
S_B(\Sigma)=\int \frac{d^6 p}{(2\pi)^6} \frac{16\pi}{p^{2}} \int dX^{\mu}\wedge dX^{\nu}\int dY^{\mu}\wedge dY^{\nu} e^{ip.(X-Y)}e^{-\epsilon |p|}.
\ee
For the wavy plane, Eq. (\ref{wavy}), we make the decomposition of $p_{\mu}$ as $(p_i,P_{i'})$, and get
\bea
S_B(\Sigma)&=&\frac{16\pi}{(2\pi)^4}\int d^4 P \int \frac{d^2 p}{(2\pi)^2} \int d^2\sigma\int d^2\sigma'\cr
&&\(1+\partial_i\phi(\sigma).\partial_i\phi(\sigma')+\delta_{ij,kl}\partial_i\phi(\sigma).\partial_k\phi(\sigma')\partial_j\phi(\sigma).\partial_l\phi(\sigma')\)\cr
&&\frac{1}{p^2+P^2}e^{ip.(\sigma-\sigma')}e^{iP.(\phi(\sigma)-\phi(\sigma'))}e^{-\epsilon |(p,P)|}.
\eea
For the scalar we have a sligthly different expression,
\bea
S_{\varphi}(\Sigma)&=&\frac{16\pi}{(2\pi)^4}\int d^4 P \int \frac{d^2 p}{(2\pi)^2} \int d^2\sigma\int d^2\sigma' \cr
&&\(1+\frac{1}{2}\((\partial\phi(\sigma))^2+(\partial\phi(\sigma'))^2\)\right.\cr
&&\left.+\frac{1}{2}\delta_{ij,kl}\(\partial_i\phi.\partial_k\phi\partial_j\phi.\partial_l\phi(\sigma)+\partial_i\phi.\partial_k\phi\partial_j\phi.\partial_l\phi(\sigma')\)\right.\cr
&&\left.+\frac{1}{4}(\partial\phi(\sigma))^2(\partial\phi(\sigma'))^2-\frac{1}{8}\((\partial \phi(\sigma))^4+(\partial\phi(\sigma'))^4\)+...\)\cr
&&\frac{1}{p^2+P^2}e^{ip.(\sigma-\sigma')}e^{iP.(\phi(\sigma)-\phi(\sigma'))}e^{-\epsilon |(p,P)|}
\eea
These two contributions we expand in powers of $\phi$ up to fourth order as $S(\Sigma)=\frac{16\pi}{(2\pi)^4}\(S_0+S_2+S_4+...\)$ where we define
\bea
S_0&=&\Gamma_0\cr
S_2&=&\frac{1}{2}\int \frac{d^2p}{(2\pi)^2} \phi(p).\phi(-p)\Gamma_2(p,\epsilon)\cr
S_4&=&-\frac{1}{8}\int \frac{d^2p_1...d^2p_4}{(2\pi)^8}(2\pi)^2\delta^2(p_1+...+p_4)\phi(p_1).\phi(p_2)\phi(p_3).\phi(p_4)\Gamma_4(p_1,...,p_4;\epsilon),
\eea
Some calculations yield the following result for the contributions from $B$ and from $\varphi$ respectively,
\bea
\Gamma_0(B)=\Gamma_0(\varphi)=(2\pi)^4\delta^2(0)H(0),
\eea
\bea
\Gamma_2(B)&=&2p^2H(p)+\frac{1}{2}G(p)\cr
\Gamma_2(\varphi)&=&\frac{1}{2}G(p)\label{result2}
\eea
and
\bea
\Gamma_4(B)&=&-\frac{1}{24}\(-2\sum_iF(p_i)+\sum_{i<j}F(p_i+p_j)\)\cr
&&-p_1.p_2\(G(p_1)+G(p_2)-G(p_1+p_3)-G(p_1+p_4)\)\cr
&&-8\delta_{ij,kl}(p_1)_i(p_2)_k(p_3)_j(p_4)_lH(p_1+p_3)\cr
\Gamma_4(\varphi)&=&-\frac{1}{24}\(-2\sum_iF(p_i)+\sum_{i<j}F(p_i+p_j)\)\cr
&&-p_1.p_2\(-G(p_3)-G(p_4)+G(p_3+p_4)\)\cr
&&-2p_1.p_2p_3.p_4H(p_1+p_2),\label{result4}
\eea
where the functions $F$, $G$ and $H$ where defined in Eq. (\ref{fns}).

\subsection{The second variational derivative}
To compute the second variational derivative of the renormalized Wilsons surface, which we write as $\<W\>=e^{-S[X]}$, we need
\bea
\frac{\delta S[X]}{\delta X(\sigma)}.\frac{\delta S[X]}{\delta X(\sigma')}-\frac{\delta^2 S[X]}{\delta X(\sigma).\delta X(\sigma')}
\eea
The first term here does not contain any delta function singularity and hence can not contribute to the surface equation. But we have to compute the second term. In momentum representation this is
\be
\frac{\delta^2 S}{\delta X^{\mu}(k)\delta X^{\mu}(k')}=\frac{\delta^2 S}{\delta X^{i}(k)\delta X^{i}(k')}+\frac{\delta^2 S}{\delta \phi^{i'}(k)\delta \phi^{i'}(k')}.
\ee
For later reference, we here let the wavy plane be a $k$-dimensional surface embedded in $D$ dimensions. The second variational derivative at lowest order comes from varying $S_2$ twice with respect to $\phi^{i'}$, and is given by
\bea
\frac{\delta^2 S_2}{\delta \phi(k).\delta \phi(k')}&=&\frac{1}{(2\pi)^2}(D-k)\delta^k(k+k')\Gamma_2(k).
\eea
At next order we get contributions from $S_2$ when varied twice with respect to $X^i$, and from $S_4$ when varied twice with respect to $\phi^{i'}$. From the condition of reparametrization invariance,
\be
0=\frac{\partial X^{\mu}(\sigma)}{\partial \sigma^i}\frac{\delta}{\delta X^{\mu}(\sigma)} = \frac{\delta}{\delta X^i(\sigma)}+\partial_i\phi^{i'}\frac{\delta}{\delta \phi^{i'}(\sigma)},
\ee
we get
\bea
\frac{\delta S}{\delta X^i(\sigma)\delta X^k(\sigma')}&=&\partial'_k\phi^{j'}(\sigma')(\partial_i\delta(\sigma-\sigma'))\frac{\delta S}{\delta\phi^{j'}(\sigma)}+\partial_i\phi^{j'}(\sigma)\partial'_k\phi^{l'}(\sigma')\frac{\delta S}{\delta \phi^{j'}(\sigma)\delta \phi^{l'}(\sigma')}
\eea
In momentum space,\footnote{We should define the Fourier transform of $\delta A/\delta \phi$ with opposite frequency to that of $\phi$ so that $\delta A = (\delta A/\delta \phi) \delta \phi$ holds in momentum space and position space alike.} 
\bea
\frac{\delta^2 S}{\delta X^i(k)\delta X^k(k')}&=&\frac{1}{(2\pi)^6}\int d^k p d^k p' p_ip'_k\cr
&&\[\delta^k(k+k'+p+p') \phi^{i'}(p)\frac{\delta S}{\delta \phi^{i'}(p')}-\phi^{i'}(p)\phi^{k'}(p')\frac{\delta^2 S}{\delta \phi^{i'}(k+p)\delta\phi^{k'}(k'+p')}\].
\eea
To second order in $\phi$ we get
\be
\frac{\delta^2 S_2}{\delta X^i(k)\delta X^i(k')}=\frac{1}{(2\pi)^4}\int \frac{d^kp d^kp'}{(2\pi)^4}  \phi(p).\phi(p')(2\pi)^2\delta^k(k+k'+p+p')s_1(p,p',k,k').
\ee
where we define
\bea
s_1(p,p',k,k')\equiv-p.p'\big(\Gamma_2(k+p)-\Gamma_2(p')\big)
\eea
Next, varying
\bea
S_4=-\frac{1}{8}\int \frac{d^2p_1...d^2p_4}{(2\pi)^6}\delta^k(p_1+...+p_4)\phi(p_1).\phi(p_2)\phi(p_3).\phi(p_4)\Gamma_4(p_1,...,p_4).
\eea
twice with respect to $\phi^{i'}$ we get
\bea
\frac{\delta^2 S_4}{\delta \phi(k).\delta \phi(k')} &=&
\frac{1}{(2\pi)^6}\int d^2 p_1 d^2 p_2\delta^2(p_1+p_2+k+k')\phi(p_1).\phi(p_2)s_2(p_1,p_2,k,k')
\eea
where we define
\bea
s_2(p_1,p_2,k,k')&\equiv&-\frac{1}{8}(D-2)\(\Gamma_4(p_1,p_2,k,k')+ \Gamma_4(k,k',p_1,p_2)\)\cr
&&-\frac{1}{8}\big(\Gamma_4(p_1,k,p_2,k')+\Gamma_4(p_1,k,k',p_2)+\Gamma_4(k,p_1,p_2,k')+\Gamma_4(k,p_1,k',p_2)\big)\cr
&&+(k\leftrightarrow k').
\eea

Restricting to the case $k'=-k$ (and $D=6$, $k=2$ dimensions) where we thus have
\bea
\frac{\delta^2 S}{\delta X^i(k).\delta X^i(-k)}&=&\frac{16\pi}{(2\pi)^4}\frac{1}{(2\pi)^4}\int \frac{d^2p}{(2\pi)^2}\phi(p).\phi(-p)s_1(p,-p,k,-k)\cr
\frac{\delta^2 S}{\delta \phi^{i'}(k).\delta \phi^{i'}(-k)}&=&\frac{16\pi}{(2\pi)^4}\frac{1}{(2\pi)^4}\((2\pi)^2\delta^2(0)\Gamma_2(k)+\int \frac{d^2p}{(2\pi)^2}\phi(p).\phi(-p)s_2(p,-p,k,-k)\)
\eea
and inserting our result Eq's (\ref{result2}) and (\ref{result4}), we get
\bea
s_2[B]&=&\frac{1}{2}\(F(p+k)-F(p)-F(k)\)\cr
&&+\(2p^2+2k^2-p.k\)G(p+k)\cr
&&-2p^2G(p)-2k^2G(k)\cr
&&+6\(p^2k^2-(p.k)^2\)H(p+k),\cr
s_2[\varphi]&=&\frac{1}{2}\(F(p+k)-F(p)-F(k)\)\cr
&&+2p^2G(k)+2k^2G(p)\cr
&&-\frac{1}{2}p.k(G(p+k)-G(p-k))+2(p.k)^2H(p+k),
\eea
symmetrized under $k\leftrightarrow -k$. We now recognize some terms in $s_2$ from the computation we did in the operator product expansion, and we expect the remaining terms to come from the remaing terms in the operator product. Explicitly we have
\bea
\frac{1}{V_3}s_1[B]&=&-\frac{3}{2}p^6\ln|p|+\frac{3}{2}p^2(p+k)^4\ln|p+k|\cr
&&-p^4k^2-\frac{1}{2}p^2k^4-2p^2(p.k)^2\cr
\frac{1}{V_3}s_2[B]&=&\frac{3}{2}p^6\ln|p| + \frac{3}{2}k^6\ln|k|\cr
&&+\(-\frac{3}{2}(p+k)^6+5p.k(p+k)^4-6(p.k)^2(p+k)^2+6p^2k^2(p+k)^2\)\ln|p+k|\cr
&&+\frac{43}{24}\((p+k)^6-p^6-k^6\)-\frac{45}{12}(p.k)\(p^2+k^2\)^4-3\(p^2k^2-(p.k)^2\)(p+k)^2,
\eea
and something very similar for $\varphi$ (displaying only the most relevant terms),
\bea
\frac{1}{V_3}s_2[\varphi]&=&-\frac{1}{2}p^6\ln|p|+...\cr
\frac{1}{V_3}s_1[\varphi]&=&\frac{1}{2}p^6\ln|p|+...
\eea
In cutoff regularization we get a harmless quadratic divergence $\epsilon^{-2}$ (corresponding to the renormalization of the string tension), but we also have $\ln \epsilon$, one for each logarithm in our result above. Thus we have in $s[B]\equiv s_1[B]+s_2[B]$ a divergence $(\frac{3}{2}\(p^6+k^6-(p+k)^6+...\)\ln\epsilon$. A rescaling of $\epsilon$ will generate a term proportional to $(p+k)^6-p^6-k^6+...$. We easily see that the $p^6\ln\epsilon$ terms cancel, which is in accordance with Eq. (\ref{central}): We see that $s[B]$ tends to zero as $k$ tends to zero. The same is true for $s[\varphi]$.

Following our recipe, we should expand\footnote{To understand that this really is the right thing to do, consider for instance $k^4\ln\(\frac{p+k}{k}\)^2$. We claim that its Fourier transform contains a piece proportional to $p^4\delta^2(\sigma-\sigma')$. To this end, consider the following integral with a test function $u(\rho)$,
\bea
&&\int d^2\rho u(\rho)\int d^2 k k^4\ln\(\frac{p+k}{k}\)^2 e^{ik.\rho}\cr
&=&\int d^2 k \tilde{u}(k)k^4\ln\(\frac{p+k}{k}\)^2 \cr
&=&\int d^2 k\((k-p)^4\tilde{u}(k-p)-k^4\tilde{u}(k)\)\ln (k^2)\cr
&=&\sum_{n=1}^{\infty} \frac{1}{n!}p^n\int d^2k k^4\tilde{u}(k)\(\frac{d}{dk}\)^4\ln(k^2).
\eea
Here $\tilde{u}(k)$ denotes the Fourier transform of $u(\sigma)$. Now picking the term for $n=4$ in this sum, we have $p^4\int d^2 k \tilde{u}(k)=p^4u(0)$. So indeed $\int d^2 k k^4\ln\(\frac{p+k}{k}\)^2 e^{ik.\rho} \sim \delta^2(\rho)+...$.} $\ln|p+k|$ in powers of $p/k$ (thus making an expansion about $k=\infty$) and then average over directions, that is replace $(p.k)^{2n}$ by $\frac{(2n-1)!!}{(2n)!!}p^{2n}k^{2n}$. That will give us a term $\sim p^6$ in $s[B]$ which should be precisely the term that we obtained in the right-hand side of the abelian surface equation, Eq (\ref{se}). Similarly we get another $p^6$-term
 in $s[\varphi]$.

The next constant term is $p^6\ln|p|$. Such a term we can never get from a rescaling $\epsilon$, not even if we had had a term $\sim p^6\ln\epsilon$. Imagine that we had got such a term somehow (though translational invariance forbids such a term) {\sl{e.g}} from a Fourier transform of a term in the operator product in the right-hand side of Eq. (\ref{second1}) such as
\bea
p^6\int d^2\sigma e^{ik.(\sigma-\sigma')}\frac{1}{|\sigma-\sigma'|^2}\sim p^6\ln(|k|\epsilon).
\eea
We could then rescale $\epsilon$ by dimensionless numbers, such as $e^2$ or $\pi$ say, which then could generate a term like $2p^6$ etcetera, but we can certainly not rescale $\epsilon$ by $e^{\ln|p|}$ for dimensional reasons. Such a rescaling would generate $p^6\ln|p|$, but is thus not permitted. The presence of a term $p^6\ln|p|$ therefore would have to come from a delta function singularity $\delta^2(\sigma-\sigma')$, simply by a dimensional analysis argument. A necessary condition for the surface equation to hold is thus that the term $p^6\ln|p|$ cancels. Here we see that it does. What we will show in the following sections is that this also happens exactly in $D=6$ dimensions when we use the AdS realization of the Wilson surface.

\section{Minimal submanifolds in AdS space}
We now move on to the realization of the Wilson surface as a minimal surface in AdS space. According to the AdS-CFT correspondence \cite{Witten}\cite{Polyakov-Gubser}\cite{Maldacena}\cite{Yang-Yee} this realization should be reliable in the large $N$ limit of $A_N$ theory. We let $M_{k+1}$ denote this minimal surface, and parametrize it as $\sigma^I\mapsto Z^{M}(\sigma)$ where $I=(0,i)$, $i=1,...,k$ and $M=(0,\mu)$, $\mu=1,...,D$. We will denote by $\tau=\sigma^0$ and $Z^M=(Y,X^{\mu})$. The submanifold $\Sigma_k$ is defined as the slice $\tau=0$ of $M_{k+1}$. The induced metrics on $M_{k+1}$ and $\Sigma_k$ are
\bea
g_{IJ}&=&G_{MN}(Z)\partial_I Z^M\partial_J Z^N
\eea
and $g_{ij}$ respectively. We define
\bea
g&\equiv &{\rm{det}}(g_{IJ})\cr
\bar{g}&\equiv &{\rm{det}}(g_{ij}).
\eea
Viewing the volume of $M_{k+1}$,
\be
A(M_{k+1})=\int_{M_{k+1}} d\tau d^{k}\sigma \sqrt{g}
\ee
as an 'action', letting $\tau$ be the 'time', we get the 'conjugate momenta' to $Z^M$ as
\be
\Pi_M=\frac{\partial \(\sqrt{g}\)}{\partial\dot{Z}^M}.
\ee
These are subject to the primary constraints
\bea
G^{MN}\Pi_M\Pi_N&=&\bar{g},\\
\Pi_M\partial_I Z^M&=&0.
\eea
The Hamilton-Jacobi equation for the minimal surface, we get by replacing $\Pi_M$ by $\frac{\delta A}{\delta Z^M}$,
\be
G^{MN}\frac{\delta A}{\delta Z^M(\sigma)}\frac{\delta A}{\delta Z^N(\sigma)}=\epsilon_{i_1\cdots i_k}G_{M_1N_1}\cdots G_{M_kN_k}\frac{\partial Z^{M_1}}{\partial \sigma^{1}}\frac{\partial Z^{N_1}}{\partial \sigma^{i_1}}\dots \frac{\partial Z^{M_k}}{\partial \sigma^{k}}\frac{\partial Z^{M_k}}{\partial \sigma^{i_k}}.\label{Ham}
\ee
Here $A=A[Z(\sigma)]$ is the area of the minimal surface that has $\Sigma$ as its boundary. This is a functional of $\Sigma$ alone. We have one Hamilton-Jacobi equation left, namely that arising from the second primary constraint. This equation is the condition of reparametrization invariance of $\Sigma_{k}$,
\be
0=\frac{\partial Z^M}{\partial \sigma^i}\frac{\delta A}{\delta Z^M(\sigma)}.
\ee

So far our derivation has been very general. But now we will consider (the euclidean version of) AdS-space. We choose coordinates such that the metric tensor is given by
\be
G_{MN}=Y^{-2}\delta_{MN},
\ee
for which the Hamilton-Jacobi equation Eq. (\ref{Ham}) becomes
\be
\sum_{\mu=1}^k\(\frac{\delta A}{\delta X^{\mu}}\)^2+\(\frac{\delta A}{\delta Y}\)^2 = \frac{\bar{g}}{Y^{2k+2}}.
\ee
Taking $Y(\sigma)=\epsilon$ constant we have,
\be
\int d^k\sigma \frac{\delta A}{\delta Y(\sigma)}=\frac{\partial A}{\partial \epsilon}
\ee
and finally get
\be
\frac{\partial A}{\partial \epsilon}=-\frac{1}{\epsilon^{k+1}}\int d^k\sigma \sqrt{\bar{g}-\epsilon^{2k+2}\(\frac{\delta A}{\delta X^{\mu}(\sigma)}\)^2}.\label{sqrt}
\ee

\subsection{The wavy plane}
We will limit ourselves to the situation where $\Sigma_{k}$ is a wavy plane of the form (\ref{wavy}) and  make an expansion in powers of $\phi^{i'}$, which we thus see as small transverse fluctuations of the plane. Here $i=1,...,k$ and $i'=k+1,...,D$. Let us denote by $\pi_{i'}=\frac{\delta A}{\delta \phi^{i'}}$. Expanding the metric in powers of $\phi$ we get
\be
\bar{g}=1+(\partial\phi)^2+\delta_{ij,kl}\partial_i\phi.\partial_k\phi\partial_j\phi.\partial_l\phi+...
\ee
and, expanding the square root in Eq. (\ref{sqrt}), we get
\bea
\frac{\partial A}{\partial \epsilon}&=&-\frac{A_0}{\epsilon^{k+1}}-\frac{1}{2}\int d^k \sigma \(\frac{1}{\epsilon^{k+1}}(\partial\phi)^2-\epsilon^{k+1}(\pi)^2\)\cr
&&+\frac{1}{8}\int d^k \sigma \(\frac{1}{\epsilon^{k+1}}\(\(\epsilon^{2k+2}(\pi)^2-(\partial\phi)^2\)^2-4\delta_{ij,kl}\partial_i\phi.\partial_k\phi\partial_j\phi.\partial_l\phi\)+4\epsilon^{k+1}(\pi.\partial\phi)^2\) +...\nonumber
\eea
where we have define $A_0=\int d^k \sigma$ and $\delta_{ij,kl}=(\delta_{ik}\delta_{jl}-\delta_{il}\delta_{jk})/2$. Undisplayed indices are contracted in obvious ways. Making the ansatz (in momentum-space)
\bea
A&=&\frac{A_0}{k\epsilon^k}+\frac{1}{2}\int d^k p \Gamma_2(p;\epsilon) \phi(p).\phi(-p)\cr
&&-\frac{1}{8}\int d^kp_1 \cdots d^kp_4 \delta^k(p_1+p_2+p_3+p_4) \Gamma_4(p_1,...,p_4;\epsilon)\phi(p_1).\phi(p_2)\phi(p_3).\phi(p_4)+...
\eea
we get 
\be
\pi_{i'}(p)=\Gamma_2(p)\phi_{i'}(-p)-\frac{1}{2}\int d^kp_1 d^kp_2 d^kp_3 \delta^k(p+p_1+p_2+p_3)\Gamma_4\phi_{i'}(p_1)\phi(p_2).\phi(p_3)+...
\ee
Inserting this into the Hamilton-Jacobi equation, using our ansatz in momentum space, we get the following equations for the coefficients functions $\Gamma_{2}$ and $\Gamma_4$,
\bea
\frac{\partial\Gamma_2}{\partial\epsilon}&=&\epsilon^{k+1}{\Gamma_2}^2-\frac{p^2}{\epsilon^{k+1}}\cr
\frac{\partial\Gamma_4}{\partial\epsilon}&=&\epsilon^{k+1}\sum_{i=1}^{4}\Gamma_2(p_i)\Gamma_4(p_1,...,p_4)-B_4(p_1,...,p_4)
\eea
where 
\bea
B_4(p_1,...,p_4)&=&C_4(p_1,...,p_4)+2D(p_1,p_2)-4D(p_1,p_3),\label{B_4}
\eea
and we define
\bea
C_4(p_1,...,p_4)&=&\epsilon^{3k+3}\Gamma_2(p_1)\Gamma_2(p_2)\Gamma_2(p_3)\Gamma_2(p_4)\cr
D(p_1,p_2)&=&-\frac{1}{2}\epsilon^{-(k+1)}p_1.p_2p_3.p_4+\epsilon^{k+1}p_1.p_2\Gamma_2(p_3)\Gamma_2(p_4).
\eea 
The equation for $\Gamma_4$ can be written
\be
\(\Gamma_4 e^G\)'=-e^G B_4
\ee
with prime denoting derivative with respect to $\epsilon$, and where
\be
G(\epsilon)=-\sum_{i=1}^4\int^{\epsilon} dY Y^{k+1}\Gamma_2(p_i;Y).
\ee
This means that we can integrate to get the solution
\be
\Gamma_4(\epsilon)=e^{-G(\epsilon)}\int_{\epsilon}^{\infty}dY e^{G(Y)} B_4(Y).
\ee
which vanishes at infinity.

\subsection{The case $k=2$}
Taking $k=1$ we get back the situation considered in \cite{Polyakov-Rychkov}. For $k=2$ (and more generally for all even $k$) we get more complicated expressions for $\Gamma_2$ and $\Gamma_4$. The solution for $\Gamma_2$ is derived in the appendix $A$. For $k=2$ we get
\be
\Gamma_2(p;\epsilon)=\frac{p^2K'(|p|\epsilon)}{\epsilon^2\(2K'(|p|\epsilon)-|p|\epsilon K(|p|\epsilon)\)},
\ee
where $K$ denotes the zeroth special Bessel function. This is the unique solution which is non-singular at infinity. Using the short-distance expansion for $K$, presented in appendix $A$, we get\footnote{Here $o(1)\supset O(\epsilon)$.} 
\bea
\Gamma_2(p;\epsilon)&=&\frac{1}{2}p^2 \epsilon^{-2}+\frac{p^4}{4}\ln \epsilon +\frac{p^4}{4}(\xi+\ln |p|)+o(1)
\eea
We may notice that the trancendent number $\xi(\equiv \gamma-\ln 2)$ can be removed by rescaling the cutoff as $\epsilon\rightarrow \epsilon e^{-\xi}$, which shows that (at least to this order) $\xi$ has no observable effect.

To get $\Gamma_4$ we first need
\bea
G(\epsilon)&=&-\sum_i \int^{\epsilon} dY Y^{3}\frac{{p_i}^2K'(|p_i|Y)}{Y^2\(2K'(|p_i|Y)-|p_i|YK(|p_i|Y)\)}\cr
&=&\sum_i\ln\(|p_i|^2{\epsilon}^2K(|p_i|{\epsilon})-2|p_i|{\epsilon}K'(|p_i|{\epsilon})\).
\eea
Here we made use of the differential equation that the Bessel function satisfies to calculate the integral. (But we did not care about the integration constant since that will anyway cancel in $\Gamma_4$). We then get
\be
\Gamma_4({\epsilon})=\frac{\int_{\epsilon}^{\infty} dY \prod_i\((|p_i|Y)^2K(|p_i|Y)-2|p_i|YK'(|p_i|Y)\) B_4(Y)}{\prod_i\((|p_i|{\epsilon})^2K(|p_i|{\epsilon})-2|p_i|{\epsilon}K'(|p_i|{\epsilon})\)}.\label{integral}
\ee
which, from (\ref{B_4}), will have the structure
\be
\Gamma_4=\frac{1}{16}\(1+\frac{\epsilon^2}{4}\sum|p_i|^2\)\(c_4(p_1,p_2,p_3,p_4)+2d(p_1,p_2)-4d(p_1,p_3)\)+o(1).\label{G}
\ee
The prefactor here came from expanding the numerator in powers of $\epsilon$.

\subsection{The second variational derivative}
With $\Gamma_4$ given as above, and with $s_1$ and $s_2$ defined as in section $3.1$, we find\footnote{We ignore the prefactor in Eq. (\ref{G}). It is easy to reinsert the contribution from this prefactor later. But this little omission will not affect our conclusions.}
\bea
s_2(p,-p,k,-k)&=&-\frac{1}{4}\big(Dc_4+(D-4)\(d(p,-p)+d(k,-k)\)-4(D-2)d(p,k)\big)+(k\leftrightarrow -k)\cr
s_1(p,-p,k,-k)&=&p^2\(\Gamma_2(p+k)-\Gamma_2(p)\).
\eea
The integrals we need to compute involve a product of four Bessel functions. We have not managed to compute such integrals exactly. Instead we expand everything in powers of $|k|$ and ignore negative powers (since we will at the end let $|k|$ tend to infinity). We illustrate the idea by concidering $c_4(p,-p,k,-k)$. That is, the contribution coming from $C_4(p,-p,k,-k)=Y^9\Gamma_2(p)^2\Gamma_2(k)^2$. Introducing the dimensionless integration variable $x=|k|Y$, we get
\bea
c_4(p,-p,k,-k)=\int_{|k|{\epsilon}}^{\infty} dx |p|^6x^5{K'\(\frac{|p|}{|k|}x\)}^2K'(x)^2.
\eea
Expanding in powers of $1/k$ we get\footnote{The condition for being permitted to use the short-distance expansion here is that the contribution to the integral, coming from the region where $x=|k|Y$ is not small, is negligible. This condition will be fulfilled here, as may be seen by considering the following schematic integral
\bea
\int_{k\epsilon}^{\infty}dx kf(x/k)g(x)
\eea
where at large $x$ we have $f,g\sim e^{-x}$ (this is roughly the large distance behavior of our Bessel function). To examine the validity of making a $1/k$-expansion we thus expand $f(x/k)=f_0(x)+f_1(x)/k+f_2(x)/k^2+...$ and assume this series to converge within some convergence radius, say at $x/k=R<\infty$. We can then split the integral as
\bea
\int_{k\epsilon}^{kR}dx \(f_0(x)+f_1(x)/k+...\)g(x) + \int_{kR}^{\infty}dx f(x/k)g(x)
\eea
By letting $k$ be sufficiently large we can use the large distance behavior of $g$ in the second integral. Assuming $f(x)<M$ for all $x>R$, we get 
\bea
\int_{kR}^{\infty}dx f(x/k)g(x)<Me^{-{kR}}
\eea
which tends to zero as $k\rightarrow \infty$ much faster than any of the terms in the $1/k$ expansion. This shows that we are permitted to expand in $1/k$.}
\bea
\int_{|k|{\epsilon}}^{\infty} dx |p|^6x^5\(\frac{|k|^2}{|p|^2}x^{-2}+\xi-\frac{1}{2}+\ln \(\frac{|p|}{|k|}x\)\)K'(x)^2+o(|k|^{-2}).
\eea
The integrals occuring in this expression we have collected in appendix $B$. Using this procedure on the other terms in $B_4$ as well, the result we get is (up to $o(|k|^{-2})$)
\bea
16c_4(p,-p,k,-k)&=&\frac{2}{3}p^4k^2+\frac{16}{75}p^6+\frac{8}{5}p^6\ln \frac{p}{k}\cr
16d(p,-p)&=&-4p^2k^2\epsilon^{-2}-4p^4k^2\ln(k\epsilon)+\(5-4\xi\)p^4k^2-p^2k^4-\frac{52}{25}p^6+\frac{12}{5}p^6\ln\frac{p}{k}\cr
16d(k,-k)&=&-4p^2k^2\epsilon^{-2}-4p^2k^4\ln(k\epsilon)+\(1-4\xi\)p^2k^4-\frac{1}{25}p^6-\frac{24}{5}p^6\ln\frac{p}{k}\cr
16d(p,k)&=&-4(p.k)^2\epsilon^{-2}-4(p.k)^2\(p^2+k^2\)\ln(k\epsilon)-4\xi(p.k)^2\(p^2+k^2\)\cr
&&+(p.k)^2k^2+\frac{13}{3}(p.k)^2p^2-\frac{119}{75}(p.k)^2\frac{p^4}{k^2}+\frac{8}{5}(p.k)^2\frac{p^4}{k^2}\ln\frac{p}{k}.
\eea
Our first observation is that we also at this order can absorb $\xi$ by the same rescaling of $\epsilon\rightarrow \epsilon e^{-\xi}$. 

We may also consider the divergent pieces here. We could have done the computation of these divergent terms for general momenta $p_i$ ($i=1,2,3,4$) simply by picking the coefficients for $Y^{-1}$ and $Y^{-3}$ directly from the integrand in $\Gamma_4$. Integrating these we get $\ln \epsilon$ and $\epsilon^{-2}$ respecctively. It is easy to see that the result above can be obtained from such a more general result by specializing to momenta $(p_i)=(p,-p,k,-k)$. We have also checked (up to fourth order in $\phi^{i'}$) that the result we get for the divergent terms is in accordance with the general result obtained in \cite{Graham-Witten},
\bea
\epsilon^{-2}\frac{1}{2}\int d^2\sigma \sqrt{\bar{g}} + \ln\epsilon \frac{1}{8}\int d^2\sigma\sqrt{\bar{g}}|\Omega|^2
\eea
where $\Omega^{\mu}$ is the Gauss curvature of the surface.

Absorbing the $\xi$'s by a rescaling of $\epsilon$ and making the averaging\footnote{One may  think this averaging is suspicious, thinking that $k_ik_j/|k|^2$ could not produce a delta function, being the Fourier transform of $\partial_i\partial_j\ln|\sigma|$. But this really must contain a delta function piece, since contracting $i$ with $j$ we get $\partial^2\ln|\sigma|\sim \delta^2(\sigma)$.}
\bea
\frac{(p.k)^2}{p^2k^2}\rightarrow \frac{1}{2}
\eea
we get 
\bea
s_1&=&\frac{1}{4}p^2(p+k)^4\ln|p+k|-\frac{1}{4}p^6\ln|p|\cr
32s_2&=&-4D(p^4k^2+p^2k^4)\ln|k|\cr
&&+4\(D-4\)p^6\ln\frac{|p|}{|k|}\cr
&&-\frac{19D+32}{15}p^6+\frac{9D+8}{3}p^4k^2+2(D-2)p^2k^4
\eea
Expanding $s_1$ about $1/k=0$, we get a constant piece $s_1=-\frac{1}{8}p^6+...$. Adding these contributions, we get
\be
s_1+s_2=-\frac{133}{480}p^6+\frac{1}{8}(D-6)p^6\ln|p|+(D-6)\frac{19}{32.15}p^6+...
\ee
Here $...$ are terms which we in Eq. (\ref{central}) denoted $o(1)$. But it should be clear by now that we consider an expansion about $k=\infty$. We notice that the inverse Fourier transform of $p^6\ln|p|\phi(p).\phi(-p)$ is an integral over the surface of a non-local expression. It therefore seems as locality gets violated unless we are in the critical dimension $D=6$.

It would of course be interesting if one also could show that the coefficient $-\frac{133}{480}$ of the $p^6$-term here also is in accordance with some right-hand side of some non-abelian surface equation in the large $N$ limit, like the one proposed in \cite{Ganor}.

Had we managed to compute $\Gamma_4$ exactly, so that we could trust our result all the way to the limit $k\rightarrow 0$, we should also have seen that the coefficient of the term $\sim p^6$ vanishes in that limit. In fact we can see that this is the case already in our result simply by interchanging the roles of $p$ and $k$. This amounts to an expansion in $1/p$. Making that exchange, we see no $p^6$-term at all, so our result is consistent with translational invariance.

\newpage
\begin{appendix}
\section{Solving the differential equation for $\Gamma_2$}
In this appendix we solve the differential equation for $\Gamma_2$,
\be
\frac{d\Gamma_2}{d{\epsilon}}={\epsilon}^{k+1}\Gamma_2^2-\frac{p^2}{{\epsilon}^{k+1}}.
\ee
Here ${\epsilon}>0$. We first define dimensionless quantities as $x=|p|{\epsilon}$, $\Gamma_2({\epsilon}) = |p|^{2+k} f_k(|p|{\epsilon})$. The differential equation we shall solve is thus
\be
{f_k}'=x^{k+1}f_k^2-x^{-k-1}.\label{DE}
\ee
First we notice that 
\be
f_k=\frac{1}{kx^k+f_{k-2}x^{2k}}
\ee
To see this, one can put this as an ansatz with $f_{k-2}$ being unspecified. Plugging this ansatz into \ref{DE} one finds that $f_{k-2}$ indeed obeys the same type of differential equation, though with $k$ replaced by $k-2$. So in this way we can reach any $f_k$ from just knowing two of them, say $f_{-2}$ and $f_{-1}$. Now it is easily seen that $f_{-1}=1$ is the only solution that is regular at infinity. So we can reach all $f_{k}$ for $k$ odd. To get the solution for $k$ even, we start with $k=-2$. Then
\be
f'_x=x^{-1}f^2-x.
\ee
Defining $f(x)=-\frac{x h'(x)}{h(x)}$ we get
\be
h''+x^{-1}h'-h=0.
\ee
This is a modified Bessel equation. 

The only solution which is regular at infinity is $h(x)=K_0(x)$ (up to a multiplicative constant, that anyway will cancel for $f(x)$). Since this will be the only Bessel function we will study in this paper, we drop the subscript, and just write $K(x)$. Its short-distance expansion is
\bea
K(x)=-\(1+\frac{x^2}{4}+\frac{x^4}{64}+...\)\ln x - \xi +\frac{x^2}{4}(1-\xi)+\(\frac{3}{128}-\frac{\xi}{64}\)x^4+...\cr
\eea
where $\xi\equiv\gamma-\ln 2$ and $\gamma=0.5772...$ is the Euler-Mascheroni constant. This expansion can be determined iteratively from the Bessel equation except for the constant $\xi$ which must be determined by requiring regularity at infinity.

The solution to the equation for $k=2$ becomes
\be
\Gamma_2(p;\epsilon)=|p|^{4}f_2(|p|\epsilon)
\ee
Here 
\bea
f_2(x)&=&\frac{1}{2x^2+f_0(x)x^4}\cr
f_0(x)&=&\frac{1}{f_{-2}(x)}\cr
f_{-2}(x)&=&-\frac{xK'(x)}{K(x)}
\eea
Thus
\be
\Gamma_2(p;\epsilon)=p^4\frac{K'(|p|\epsilon)}{2p^2\epsilon^2K'(|p|\epsilon)-|p|^3\epsilon^3K(|p|\epsilon)}
\ee

\section{Integrals in $\Gamma_4$}
We find it very fruitful to write the basic integrals we need as functions of a parameter $\lambda$ that we introduce as follows, 
\bea
f_p(\lambda)&\equiv &\int dx x^pK(\lambda x)^2\cr
g_p(\lambda)&\equiv &\int dx x^pK'(\lambda x)^2
\eea
Here $K$ is as in appendix $A$, and $p=1,3,5,7,...$. Using the differential relation $K''(\lambda x)=K(\lambda x)-(\lambda x)^{-1}K'(\lambda x)$ it is then easy to get at the following relations,
\bea
(p+1)f_p-(p-1)g_p&=&x^{p+1}\(K^2-K'^2\)\cr
f_p&=&\frac{1}{2p}\(x^{p+1}\(K^2-K'^2\)+\frac{p-1}{2}\(\ddot{f}_{p-2}+\lambda^{-1}\dot{f}_{p-2}\)\)
\eea
where $\cdot\equiv \frac{d}{d\lambda}$ and $'$ denotes the usual derivative, with respect to the argument of the functions (in this case $\lambda x$). The latter relation is a recursion relation which enable us to compute the integrals $f_{p}$ for $p$ odd. Happily, integrals with even $p$ do not show up in this paper. Also notice that the initial `value' $f_1=\frac{1}{2}x^{2}\(K^2-K'^2\)$ actually gets determined from the recursion relation. We also get $g_{-1}=\frac{1}{2}\(K^2-K'^2\)$. But we must supply an additional relation for $g_1$, for instance
\bea
g_1+f_1&=&\lambda^{-1}xKK'.
\eea

We will only be interested in the short distance behavior of $f_p(\lambda,x)$ when $x$ is small, which we now should define as 
\bea
-f_p(x,\lambda)=\int_x^{\infty}dy y^pK(\lambda y)^2
\eea
and similarly for $g_p$. We will actually only need the constant piece, and are not interested in the terms in $o(1)\supset O(x)$. Using the short-distance expansions $K(y)=-\ln y-\xi+o(1)$ and $K'(y)=-y^{-1}+o(1)$, and the far distance behavior, $x^p K(x)\rightarrow 0$ as $x\rightarrow\infty$ for any $p>0$, and applying our recursion relation, we then get for the first few orders 
\bea
f_1(x,\lambda)&=&-\frac{1}{2}\lambda^{-2}+o(1)\cr
f_3(x,\lambda)&=&-\frac{1}{3}\lambda^{-4}+o(1)\cr
f_5(x,\lambda)&=&-\frac{16}{15}\lambda^{-6}+o(1)
\eea
and 
\bea
g_1(x,\lambda)&=&\(\ln (\lambda x)+\xi´+\frac{1}{2}\)\lambda^{-2}+o(1)\cr
g_3(x,\lambda)&=&-\frac{2}{3}\lambda^{-4}+o(1)\cr
g_5(x,\lambda)&=&-\frac{8}{5}\lambda^{-6}+o(1)
\eea

We will also need the integrals
\bea
h_p(\lambda)&\equiv&\int dx x^p(\ln x) K(\lambda x)^2\cr
k_p(\lambda)&\equiv&\int dx x^p(\ln x) K'(\lambda x)^2
\eea
which are subject to the relations
\bea
(p+1)h_{p}-(p-1)k_{p}&=&x^{p+1}\ln x \(K^2-K'^2\)-\(f_p-g_p\)\cr
h_p&=&\frac{1}{2p}\(x^{p+1}\ln x (K^2-K'^2)+\frac{p-1}{2}\(\ddot{h}_{p-2}+\lambda^{-1}\dot{h}_{p-2}\)-\(f_p-g_p\)\)
\eea
and again we have to supplement a relation, such as
\bea
h_1+k_1=\lambda^{-1}\(x(\ln x)KK'-\frac{1}{2}\lambda^{-1}K^2\).
\eea
From these relations we derive the following short-distance behaviors
\bea
h_1(x,\lambda)&=&\frac{1}{2}\lambda^{-2}\(1+\xi+\ln \lambda\)+o(1)\cr
h_3(x,\lambda)&=&\frac{1}{3}\lambda^{-4}\(-\frac{1}{6}+\xi+\ln \lambda\)+o(1)\cr
h_5(x,\lambda)&=&\frac{4}{15}\lambda^{-6}\(-\frac{43}{225}+4(\xi+\ln\lambda)\)+o(1)
\eea
and 
\bea
k_1(x,\lambda)&=&\lambda^{-2}(...)\cr
k_3(x,\lambda)&=&\frac{2}{3}\lambda^{-4}\(\xi+\ln\lambda\)+\frac{1}{18}\lambda^{-4}+o(1)\cr
k_5(x,\lambda)&=&\frac{8}{5}\lambda^{-6}\(\xi+\ln\lambda\)-\frac{76}{75}\lambda^{-6}+o(1)
\eea
We here collect all the integrals that occur in $\Gamma_4$:
\bea
\int_x^{\infty} dy y^{-1}(2K'-yK)^2&=&2x^{-2}+2(\ln x+\xi)-\frac{1}{2}+o(1)\cr
\int_x^{\infty} dy y(2K'-yK)^2&=&-4(\ln x+\xi)+\frac{13}{3}+o(1)\cr
\int_x^{\infty} dy y^3(2K'-yK)^2 &=& \frac{32}{5}+o(1)\cr
\int_x^{\infty} dy yK'^2&=&-\(\ln y+\xi+\frac{1}{2}\)+o(1)\cr
\int_x^{\infty} dy y^3K'^2&=&\frac{2}{3}+o(1)\cr
\int_x^{\infty} dy y^5K'^2&=&\frac{8}{5}+o(1)\cr
\int_x^{\infty} dy \ln y y^3(yK-2K')^2&=&\frac{124}{75}-\frac{32}{5}\xi+o(1)\cr
\int_x^{\infty} dy \ln y y^5K'^2&=&\frac{76}{75}-\frac{8}{5}\xi+o(1).
\eea

\section{Conformal invariance}
We here generalize the proof of conformal invariance of the loop operator given in \cite{Polyakov-Rychkov}, to surface operators acting on $k$-dimensional submanifolds embedded in ${\bf{R}}^D$ parametrized as $\sigma_i\mapsto X_{\mu}(\sigma)$. To this end, consider a conformal transformation $x_{\mu}\rightarrow \tilde{x}_{\mu}(x)$ for which
\be
\partial_{\mu}\tilde{x}_{\lambda}\partial_{\mu}\tilde{x}_{\sigma}=\Omega(x)\delta_{\lambda\sigma}
\ee
for some function $\Omega(x)$. This induces a natural transformation $X\mapsto \tilde{X}$ of the coordinates of the submanifold. Introducing the notations $\tilde{W}[X]\equiv W[\tilde{X}(X)]$, and $\partial_{\mu}\tilde{x}_{\nu}(X(\sigma))\equiv \frac{\partial\tilde{x}^{\nu}}{\partial x^{\mu}}\Big|_{x=X(\sigma)}$, repeated use of the chain rule yields \be
\frac{\delta^2 \tilde{W}[X]}{\delta X_{\mu}(\sigma)\delta X_{\mu}(\sigma')}
=\delta^{k}(\sigma-\sigma')\partial_{\mu}\partial_{\mu}\tilde{x}_{\lambda}(X(\sigma))\frac{\delta {W}[\tilde{X}]}{\delta \tilde{X}_{\lambda}(\sigma)}+\partial_{\mu}\tilde{x}_{\lambda}(X(\sigma'))\partial_{\mu}\tilde{x}_{\sigma}(X(\sigma))\frac{\delta^2 {W}[\tilde{X}]}{\delta \tilde{X}_{\lambda}(\sigma')\delta \tilde{X}_{\sigma}(\sigma)}.
\ee
We have the following relation between area derivative \cite{Migdal}\cite{Polyakov} $N_{\lambda\nu_1...\nu_k}\equiv \frac{\delta{\tilde{W}}}{\delta \sigma_{\lambda\nu_1...\nu_k}}$ (here $d\sigma_{\lambda\nu_1...\nu_k}$ denotes the antisymmetric area element in directions corresponding to the indices) and functional derivative:
\be
\frac{\delta {W}}{\delta {X}_{\lambda}(\sigma)}=N_{\lambda\nu_1...\nu_k}\partial_1 {X}_{\nu_1}(\sigma)\cdots \partial_k {X}_{\nu_k}(\sigma).
\ee
Here all we need to know about the tensor $N$ is that it is completely antisymmetric. This fact follows from the condition of reparametrization invariance $\partial_i X_{\mu}(\sigma)\frac{\delta}{\delta X_{\mu}(\sigma)}=0$ of the surface. Defining the $k$-submanifold operator $L$ in a similar way as in Eq. (\ref{L}), we get 
\bea
L(\sigma)\tilde{W}[X]-\Omega(X)L(\sigma)W[X]&=&\Bigg[\(\partial^2 \tilde{x}_{\lambda}\) \(\partial_{\mu_1}\tilde{x}_{\nu_1}\cdots \partial_{\mu_k}\tilde{x}_{\nu_k}\)\cr
&&-\sum_{i=1}^{k}\(\partial_{\mu_i}\partial_{\mu}\tilde{x}_{\lambda}\)\(\partial_{\mu}\tilde{x}_{\nu_i}\)\(\partial_{\mu_1}\tilde{x}_{\nu_1}\cdots \widehat{\partial_{\mu_i}\tilde{x}_{\nu_i}}\cdots \partial_{\mu_k}\tilde{x}_{\nu_k}\)\Bigg]\cr
&&\times\partial_{1}X_{\mu_1}\cdots\partial_k X_{\mu_k} N_{\lambda\nu_1...\nu_k}.
\eea
where the hat means that this factor is removed. Considering the special conformal transformation $\tilde{x}_{\mu}=x_{\mu}/|x|^2$, this becomes
\bea
&=&\Bigg[(4-2D) \frac{x_{\lambda}}{|x|^{2k+2}} \delta_{\mu_1\nu_1}\cdots \delta_{\mu_k\sigma_k}-\sum_{i=1}^{k}\(\frac{1}{|x|^4}\(-2\delta_{\mu_i\mu}x_{\lambda}-2\delta_{\mu\lambda}x_{\mu_i}-2\delta_{\lambda\mu_i}x_{\mu}\)+\frac{1}{|x|^6}8x_{\mu_i \mu \lambda}\)\cr
&&\times\prod_{i=1}^{k}\(\delta_{\mu\nu_i}-2\frac{x_{\mu}x_{\nu_i}}{|x|^2}\)\Bigg]\partial_{1}X_{\mu_1}\cdots\partial_k X_{\mu_k} N_{\lambda\nu_1...\nu_k}.\cr
&=&(4-2D+4k)\frac{x_{\lambda}}{|x|^{2k+6}}\delta_{\mu_1\nu_1}\cdots \delta_{\mu_k \nu_k}\partial_1 X_{\mu_1}\cdots\partial_k X_{\mu_k} N_{\lambda\nu_1...\nu_k},
\eea
that is, we have found that the surface operator $L$ is conformally invariant in the sense given above, only in $D=2k+2$ dimensions.

Now consider the abelian surface equation, 
\bea
L(\sigma)W[X]=J[X]W[X]
\eea
where $J[X]$ is given by Eq.(\ref{J}) (suppressing the marked point $X(\sigma)$).
Assuming we are in the critical dimension $D=6$ we get under a conformal transformation, according to the general result above,
\bea
\Omega(X(\sigma))L(\sigma)W[X]=J[\tilde{X}]\tilde{W}[X].
\eea
Formally, or for the unrenormalized Wilson surface, we have $\tilde{W}[X]=W[X]$. We then would need to have $J[X]=\Omega(X(\sigma))^{-1}J[\tilde{X}(X)]$. Indeed we have
\bea
\delta^D(\tilde{x})=\Omega(x)^{-D/2}\delta^D(x),
\eea
and $\delta^D(X(\sigma)-X(\sigma'))$ transforms in the same way as $\delta^D(x)$. Now take $D=6$. Then we have also (formally) an additional factor $\Omega(X)^2$ coming from $\dot{\tilde{X}}_{[\mu}\tilde{X}'_{\nu]}\dot{\tilde{X}_{\mu}}\tilde{X}'_{\nu}=\Omega(X)^2\dot{X}_{[\mu}X'_{\nu]}\dot{X}_{\mu}X'_{\nu}$ (formally we could put $\sigma=\sigma'$ due to the delta function). We thus see that the abelian surface equation is formally conformally invariant. But this holds just formally. It is not true in the renormalized theory where we have to take into account the conformal anomaly of $W[X]$. We were also cheating in our derivation of the transformation law of $J[X]$. It would be interesting to see whether the surface equation is conformally invariant in quantum theory, and whether the conformal anomaly of the abelian Wilson surface can be derived directly from this surface equation.

\end{appendix}

\end{document}